\documentclass[a4paper,11pt]{article}
\usepackage{jcappub} % for details on the use of the package, please see the JINST-author-manual
\usepackage{lineno}
\usepackage{graphicx,amssymb,amsmath,xcolor,graphicx}
\usepackage[normalem]{ulem}
\arxivnumber{2311.14628} % Only if you have one
\title{A nearby source of ultra-high energy cosmic rays}

\author{Mikhail Yu. Kuznetsov}
\affiliation{Institute for Nuclear Research of the Russian Academy of Sciences,\\
60th October Anniversary Prospect 7a, Moscow 117312, Russia}

\emailAdd{mkuzn@inr.ac.ru}

\abstract{Recently the Telescope Array collaboration reported an observation of cosmic ray event with very high energy $244$~EeV ($2.44 \times 10^{20}$~eV). Importantly, the event is hard to correlate with the matter distribution in the local Universe, even after taking into account deflections in magnetic fields. This implies that the event is likely a nucleus with a large charge. An attenuation length of the nucleus of such a high energy in intergalactic space is quite small, therefore its source should be relatively close to our Galaxy. Using these arguments we derive a new upper bound on a distance to the closest ultra-high energy cosmic ray (UHECR) source and a lower bound on the UHECR source number density  in general.  The distance to the closest source should not exceed 5~Mpc at 95\%~C.L. and the $95\%$~C.L. lower-bound on the sources number density is $\rho > 1.0 \times 10^{-4}$~Mpc$^{-3}$. The number density of UHECR sources emitting heavy nuclei is constrained for the first time.}

\begin{document}
\maketitle
\flushbottom

\section{Introduction}
\label{sec:intro}
% General introduction
Ultra-high energy cosmic rays are charged particles with the energy higher than $E \gtrsim 10^{18}$~eV, that are reaching the Earth from space.
The question of UHECR origin is one of the most interesting puzzles in modern astrophysics. Despite decades of observations with various experiments and extensive theoretical modeling the nature of UHECR sources is still not clear.
An uncertain mass (and charge) composition of these particles together with an uncertain galactic and extragalactic magnetic fields makes the identification of their sources an extremely hard task.
There were a number of studies trying to identify UHECR sources~\cite{PierreAuger:2018qvk, PierreAuger:2022axr, PierreAuger:2023mvf} or at least to constrain their properties~\cite{Hillas:1984ijl, Ptitsyna:2008zs, Lang:2020wrr, Sotirov:2022uqk, Kachelriess:2022phl}.

% Introduction on TA event and constraints idea
Recently the Telescope Array experiment has reported the observation of an exceptional UHECR event with energy $E_0 = 244 \pm 29\, ({\rm stat.})\, ^{+51}_{-76}\,({\rm syst.})$~EeV~\cite{doi:10.1126/science.abo5095}.
The event direction was measured with a relatively small uncertainty of $0.8^\circ$. Unfortunately, its primary type was not determined certainly, only a gamma-ray primary was rejected with $\sim 3.8 \sigma$ confidence level.
In this letter we argue that the type of the particle could be estimated with a reasonable confidence by comparing its arrival direction with a conservatively expected UHECR flux distribution: a particle with a larger charge should deflect from the expected sources stronger than the particle with a smaller charge.
It was shown in the TA paper that the event arrival direction is pointing out towards the middle of the local LSS void. Even after taking into account deflections of the particle by the galactic magnetic field its correlation with expected sources remains elusive. This implies that the particle is unlikely to be a proton.
In what follows we quantify these arguments and find that the particle is likely a heavy nucleus.
While we do not aim to find or constrain the source of the particular highest energy particle, it is reasonable to assume that it is originating from one and the same population of sources as all the UHECR flux at sufficiently high energies.
Thus, knowing the type of the highest energy particle  allows us to place a new upper bound on the distance to the closest  source of UHECR flux in general. In turn, this leads us to  a new lower limit on the number density of UHECR sources that are tracing LSS and emitting heavy nuclei.
These limits constrain yet uncovered region of UHECR source parameter space and complement the similar constraints derived for sources emitting light or intermediate particles~\cite{PierreAuger:2013waq}.

The paper is organised as follows.
First, we describe how the sky map of the expected UHECR flux is built.
Then we consider various scenarios where the detected particle is a proton and show that all of them are unable to explain the observed lack of correlation between the particle and the expected sources.
Then we compute the constraints on the distance to the closest UHECR source and on the number density of these sources  in general. 
The procedure is organized in three steps.
First, we consider various sky maps that we expect to observe if UHECR sources are emitting nuclei and constrain the charge of the detected particle from the correlation analysis.
Second, we simulate in detail the propagation of the flux of the respective nuclei and find the distance from where the most part of expected flux is coming, that we interpret as a bound on a distance to the closest UHECR source.
Third, using this distance we compute the constraint on the average number density of the UHECR sources in the Universe.
Finally, we discuss what classes of potential UHECR sources are disfavored by this bound.

\begin{figure*}
\begin{center}
 \includegraphics[width=0.49\columnwidth]{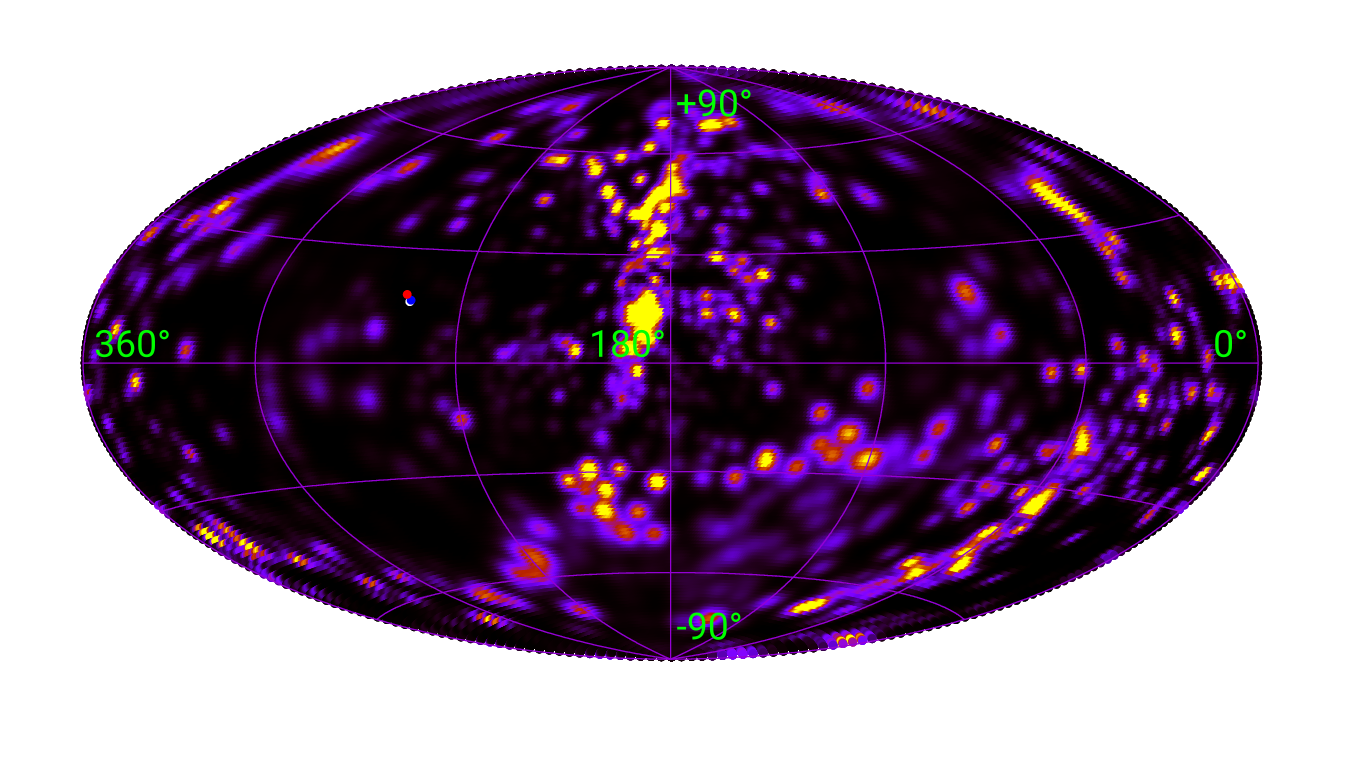}
 \includegraphics[width=0.49\columnwidth]{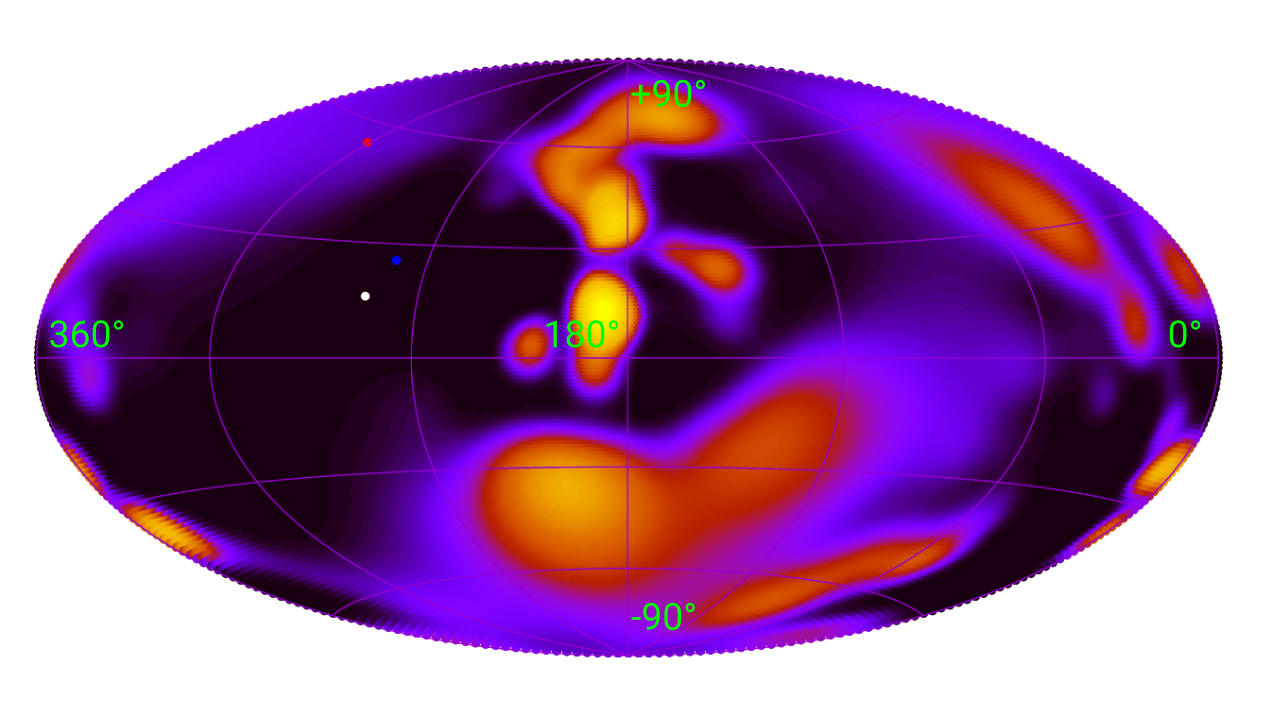}
\caption{
\label{fig:Fujii-p-Fe_original}
Correlation of the TA highest energy event with the expected UHECR flux maps.
Source smearing due to non-uniform random GMF is assumed. Deflections in regular GMF is accounted for with backtracking of the detected event direction. The detected direction is shown with the white dot, deflection in PT11 model --- with the blue dot and deflections in JF12 model --- with the red dot. No deflections in EGMF is assumed. The uncertainty of event direction measurement is smaller than the size of the dot. Maps are in equatorial coordinates (J2000).
{\bf Left panel:} the map for proton event with energy 244~EeV {\bf Right panel:} the map for iron event with energy  220~EeV.  }
\end{center}
\end{figure*}

\section{UHECR flux model}
\label{sec:flux_model}
% Flux models
To estimate the degree of the UHECR primary deflection from its expected sources and to constrain its charge we build a conservative model of the expected UHECR flux.
We assume that the sources of UHECR are tracing the distribution of luminous matter in the local Universe.
Technically, we build the UHECR flux model in the same way as in Ref.~\cite{Kuznetsov:2020hso}. In particular we assume that each galaxy in a complete volume-limited sample has the same injected UHECR flux.
We use a highly complete flux-limited sample of galaxies obtained from the 2MRS galaxy catalog~\cite{Huchra:2011ii}.
This sample excludes galaxies with magnitudes greater than 12.5 and distances less than 5 Mpc or greater than 250 Mpc.
The sources that are closer than 5 Mpc are omitted in order to treat the whole flux model on a statistical basis.
To correct the observational selection bias inherent in a flux-limited sample, we assign progressively larger fluxes to more distant galaxies.
We assume that sources beyond 250 Mpc are uniformly distributed with the same average density as those at that distance.
Their contribution is included as a properly normalized fraction of isotropic events.
The details of catalog construction can be found in Ref.~\cite{Koers:2009pd}.
We derive the injection spectrum for each nucleus from the separate fit to the Telescope Array and Auger observed spectra~\cite{diMatteo:2017dtg, Kuznetsov:2023xly}.
As a result we get power-law injection without cutoff and with slopes -2.55 and -1.89 for protons and iron, respectively.
While the assumption of no-cutoff is not realistic, it is conservative for constraining the correlation between event and sources, which is our goal.

The deflections in magnetic fields are computed taking into the account the primary particle charge Z and its energy E. We simulate the deflections in the regular galactic magnetic field (GMF) using the backtracking method either with Pshirkov-Tinyakov (PT11) GMF model~\cite{Pshirkov:2011um} or Jansson-Farrar (JF12) model~\cite{Jansson:2012pc}. The deflections in the random GMF is simulated  using a galactic-latitude-dependent Gaussian smearing that follows the data-driven formula of Ref.~\cite{Pshirkov:2013wka}~\footnote{ Note, that the mentioned relation is based on a phenomenological fit for Faraday rotation measures of extragalactic sources, which is independent of any assumption about the coherence length of random GMF.}.
The UHECR deflections in EGMF can be non-negligible and even higher than those in GMF.
Therefore for the most conservative propagation scenario we adapt the largest observationally allowed value of $B_{\rm EGMF} = 1.7$~nG with $\lambda_{\rm EGMF} = 1$~Mpc~\cite{Pshirkov:2015tua}. We implement the effect of this field as an additional direction-independent smearing of the expected UHECR flux map.
For convenience we attribute the smearing due to the random magnetic fields to source maps, but the deflection due to regular magnetic field --- to event direction.

\section{UHECR correlation with LSS for various primary types}
\label{sec:corr_tests}
First, we test the expected maps for proton and iron primaries
without the EGMF effect.
To be more accurate, for the proton primary we use the central value of detected energy $E_0 = 244$~EeV, while for the iron primary we use $\tilde{E_0} = 220$~EeV --- the value reduced by $10\%$, which is the systematic bias of the experiment energy reconstruction for the iron nucleus~\cite{doi:10.1126/science.abo5095}.
In Fig.~\ref{fig:Fujii-p-Fe_original} we confront these maps with positions of proton and iron events backtracked according to the mentioned regular GMF models.
Considering the expected flux maps as normalized probability density maps of UHECR arrival directions we found a p-value for proton primary to arrive from the given direction to be less than $1\%$ for both regular GMF models~\footnote{Just before the present paper was finished, a set of new regular GMF models was released~\cite{Unger:2023lob}. Using results from another recent paper of the same authors~\cite{Unger:2023hnu} we estimated the magnitude of UHECR deflection in new models at the position of the highest energy TA event to be maximum a factor 1.3 of the deflection in JF12 GMF model. Using this estimation with our expected flux maps we found that the p-value for proton is less than $1\%$ for new GMF models too.}. The similar computation for iron primary gives $p \sim 16\%$ for JF regular GMF model (while still $p < 1\%$ for PT regular GMF model), allowing for the conclusion that the primary can be an iron nucleus.

To make this conclusion more robust we need to consider the possible effect of EGMF and an uncertainty of the detected event energy. We set the energy of the detected particle to the lower edge of its uncertainties, assuming $E_{\rm real} = E_0 - 2 \sigma_{\rm stat.} - \sigma_{\rm syst.} = 135\, {\rm EeV}$. For the estimation of the primary proton deflection in the EGMF we conservatively assume the distance to the source $L=250$~Mpc (the upper edge of our catalog), that yields $5.2^\circ$ smearing at 135 EeV.
Even with these conservative assumptions
it is hard to correlate the event with the sources in LSS: 
the p-values to get a proton primary from the given directions are still less than $1\%$ for both regular GMF models.

\begin{figure}
 \includegraphics[width=0.99\columnwidth]{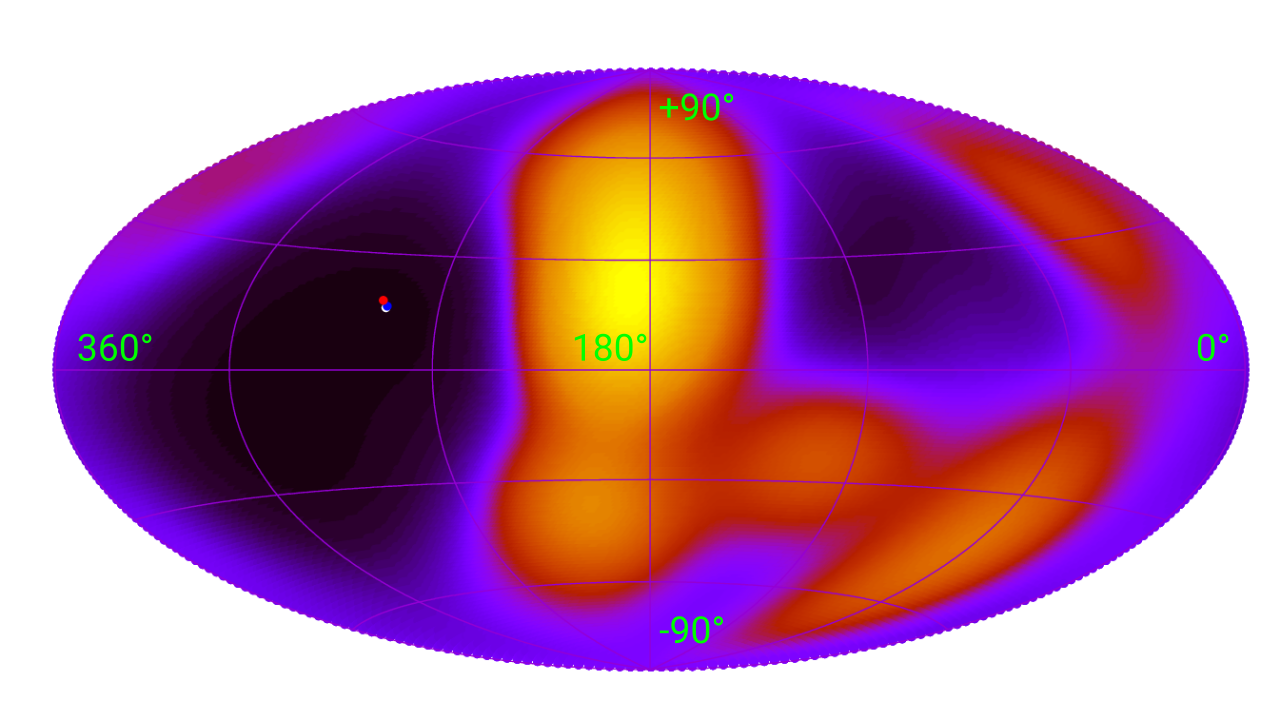}
\caption{
\label{fig:map_corr}
Correlation of the TA highest energy event with the expected UHECR flux map for the injected iron nucleus and detected secondary proton at $E > 168$~EeV.
The flux model is the same as in Fig.~\ref{fig:Fujii-p-Fe_original} plus the uniform smearing of the sources in the maximum allowed EGMF for iron charge: $Z=26$ (see text). Deflections in the regular and random GMF is for proton charge. The map is in equatorial coordinates (J2000).
}
\end{figure}

One can imagine a more complicated scenario, where the source emits a nucleus but we detect a secondary proton that was spallated from the primary particle due to its interaction with a background photon field upon the propagation through intergalactic medium.
It this case one can expect that the nucleus would be deflected strongly by EGMF and then emit a proton that would propagate to a larger distance than the primary, because proton attenuation length is larger in the energy range we consider.
However, the computations show that this is not the case in our setup, even if we conservatively allow the primary iron nucleus to be deflect by the EGMF along its full path to the Earth, and then emit a proton that is deflected by GMF. Indeed, in the conservative case of injected energy $E_{inj} = 10^{22}$~eV and the detected energy corresponding to the lower edge on the energy systematic uncertainty: $E_0 = 1.68 \cdot 10^{20}$~eV) the full path is $14.0$~Mpc and the deflection in EGMF for iron nucleus is $24^\circ$. The resulting map is shown in Fig.~\ref{fig:map_corr}, one can see that the p-value to have a particle from this direction at the resulting sky map would be again less than $1\%$.

\section{Constraints on the distance to the closest UHECR source}
\label{sec:constr_L}

\begin{figure}
\includegraphics[width=0.99\columnwidth]{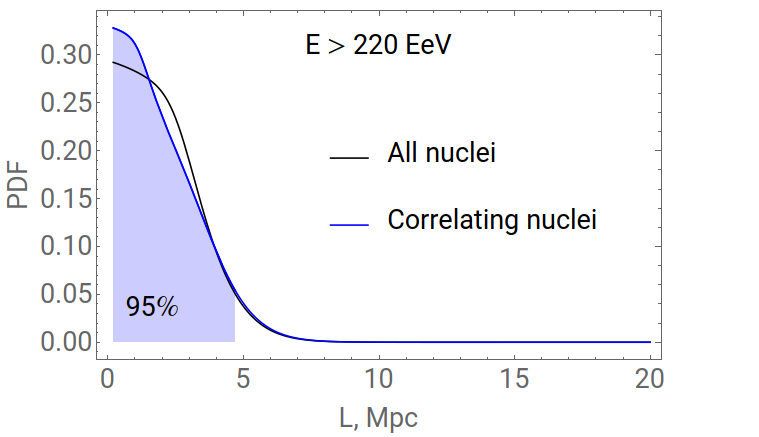}
\caption{
\label{fig:flux}
The relative integral flux of UHECR originating from within distance $L$ for energies  $\tilde{E_0} > 220$~EeV  and injected iron nucleus. The uniform spatial distribution of sources is assumed. 
Flux that includes all the secondary cascade is shown with the black line and the flux that includes only secondaries correlating with LSS --- with the blue line.
The distance  $L_0 = 4.7$~Mpc  from where the $95\%$ of the correlating flux is coming is shown with a blue shading.
}
\end{figure}

Taking into account the above arguments we further consider the TA highest energy event to be a secondary product of an injected iron nucleus primary.
We need to note that the primary light or intermediate nucleus cannot be excluded, also.
However, we do not consider these possibilities, as the attenuation lengths of light nuclei are smaller than that of iron, so that the respective upper bound for $L$ would be conservative for iron primary for $E > \tilde{E_0}$.
To constrain the distance to the source of the discussed iron nucleus we use two step strategy. First, we consider the UHECR flux maps expected for various secondaries of iron primary.
To place the 95\%~C.L. upper limits we found the minimal nucleus charge $Z_0$ for which the p-value of the respective map is reaching at least $5\%$, so that all nuclei in the cascade with $E>\tilde{E_0}$ and $Z>Z_0$ are fulfilling the correlation with the sources in that sense.
The secondary  silicon ($Z_0 = 14$)  is the minimal correlating nucleus for $\tilde{E_0} = 220$~EeV.
Then we use the TransportCR code~\cite{Kalashev:2014xna} to simulate the propagation of the injected iron nuclei, assuming the uniform spatial distribution of sources and taking into account all secondaries.  The extragalactic background light model of Ref.~\cite{Franceschini:2017iwq} is used~\footnote{ The usage of the minimal model of Ref.~\cite{Kneiske:2010pt} as an alternative is affecting the result by less than $1\%$.}.
We compute the flux $F_0$ of all nuclei with $E>\tilde{E_0}$ and $Z>Z_0$ reaching the Earth from the sources within radius $L$ and the similar flux for $L \rightarrow \infty$.
Their ratio is interpreted as a probability density to have the source not farther than $L$.
In Fig.~\ref{fig:flux} we show the resulting fluxes.
We interpret the distance $L_0$ from where the $95\%$ of the total possible flux originating as a 95\%~C.L. upper bound for the distance to the  closest  source of  UHECR flux. For $\tilde{E_0} = 220$~EeV we have $L_0 < 4.7$~Mpc. 
However, conservatively our bound is $L_0 < 5$~Mpc, as the LSS catalog for which our expected flux maps were computed does not contain sources closer than 5~Mpc, so that we cannot claim the lack of correlation with the possible closer sources.  We need to stress, that the limit $L_0 < 5$~Mpc is conservative in a sense that the possible correlation of the event with closer sources can only make this limit stronger but not weaker.
Repeating the same analysis for $E > 168$~EeV, which is the lower edge on the systematic uncertainty of the TA event energy, we get $L_0 < 13.5$~Mpc.
We want to stress, that these results do not depend on the maximum energy of particles injected in the sources.

\section{Constraints on the UHECR source number density}
\label{sec:constr_rho}

Using these distances one can place a lower bound on a number density of UHECR sources $\rho$.
To make the bounds not only meaningful for our local extragalactic neighborhood but universal we assume that UHECR sources are distributed in space according to the Poisson distribution:
\begin{equation}
    P(\rho, N) = \frac{e^{-\rho V} (\rho V)^N}{N!}
\end{equation}
where $V$ is a volume inside a sphere of radius $L_0$, and $N$ is a number of sources.
For a given source number density $\rho$ we generate a large number of 3D source samples and find a fraction of realizations with at least one source in a volume $V$.
Value $\rho_0$ for which this fraction is no more than $5\%$ is the $95\%$~C.L. lower bound for the UHECR source number density. This procedure yields $\rho > 1.0 \cdot 10^{-4}$~Mpc$^{-3}$ for $L_0 = 5$~Mpc and respective $\tilde{E_0} = 220$~EeV. 
To properly compare our result with that of the Pierre Auger experiment~\cite{PierreAuger:2013waq} we compute the similar systematic uncertainty in the source number density due to the systematics in event energy estimation.
The lower edge of the event energy systematic uncertainty is $168$~EeV. This translates to the following lower bounds of source number density: $\rho > 5.1 \cdot 10^{-6}$~Mpc$^{-3}$.
 We need to note that these bounds are definitely applicable from highest UHECR energies down, while the events are still originating from the same source population. Moreover, if at some lower energy the extragalactic UHECR are originating from another source population, we expect these bounds to be applicable too, as the sources of lower energy particles are expected to be more common than those of higher energy particles.

\section{Discussion}
\label{sec:discussion}

\begin{figure}
  \includegraphics[width=0.99\columnwidth]{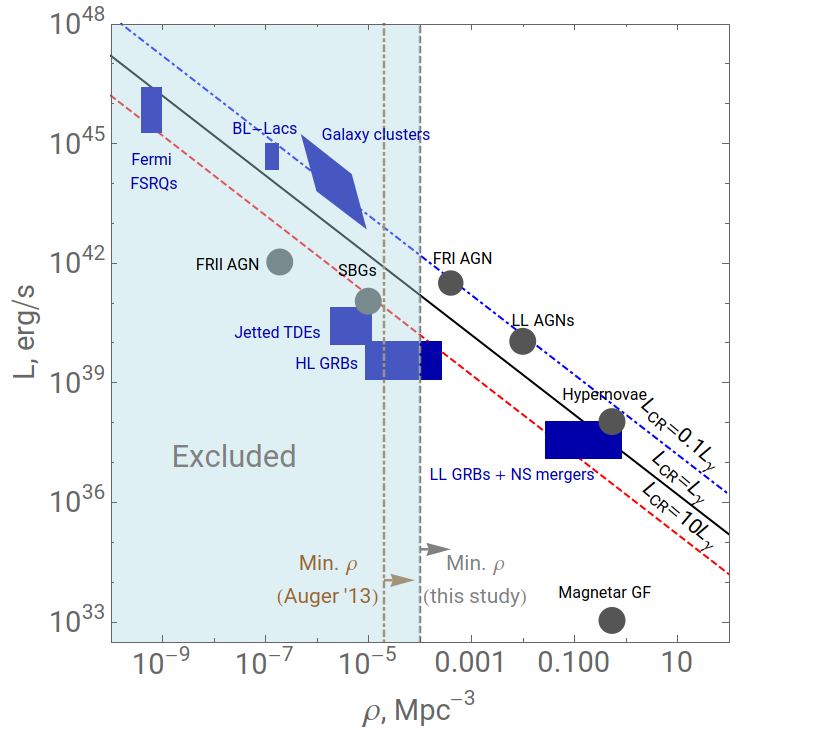}
\caption{
\label{fig:L_vs_rho}
Plot of effective source luminosity versus source number density. Several steady and transient source classes are shown. For transient sources the lifetime is set to $3 \cdot 10^5$~yr. The black solid line shows the best-fit for total UHECR luminosity density $L_{CR} = 5 \cdot 10^{44}\, {\rm erg\, Mpc^{-3} yr^{-1}}$ from Ref.~\cite{Aab:2016zth}.
The figure is adapted from Ref.~\cite{AlvesBatista:2019tlv}.
The dashed vertical line is the lower bound for the source number density from the present study. The dot-dashed vertical line is the bound from Ref.~\cite{PierreAuger:2013waq} for the $30^\circ$ deflections and sources in LSS.
}
\end{figure}
We would like to note that our constraints are complementing those of Pierre Auger collaboration~\cite{PierreAuger:2013waq}. Those constraints were set for particles with deflections not larger than $30^\circ$, which implies light or intermediate composition. Indeed, the deflection of $100$~EeV proton in the galactic magnetic field is approximately $3^\circ$, when estimated as a median over all directions in the sky in either JF12 or PT11 GMF model~\cite{IceCube:2022osb}. Thus, Auger result obtained for $E > 80$~EeV CRs should be attributed to the {\it observed} particles with $Z \lesssim 8$. This implies the respective injected particles are likely lighter than silicon nuclei --- simulations show that in the case of injection of silicon and heavier nuclei these primaries dominate the observed flux at $E \simeq 80$~EeV, unless the rigidity is higher than $80$~EV, and the observed flux is dominated by secondary protons.
In contrast, our present constraints are viable for sources emitting heavy particles, up to iron nuclei. We need to note that according to the recent TA studies the UHECR flux at energies higher than $100$~EeV consists of just heavy nuclei~\cite{Kuznetsov:2023xly}.

In Fig.~\ref{fig:L_vs_rho} we show our constraints on a source number density versus source luminosity plot, along with the constraints of Auger~\cite{PierreAuger:2013waq}. The regions of parameter space covered by various source models follow the results of Ref.~\cite{AlvesBatista:2019tlv}. One can see that the present constraints improve older ones by five times: if we consider intermediate or heavy composition, that is suggested by data at highest energies~\cite{Aab:2016zth, PierreAuger:2022atd, Kuznetsov:2023xly}. Therefore, galaxy clusters, starburst galaxies and jetted tidal disruption events appear to be disfavored as main sources of UHECR~\footnote{We need to stress, that this conclusion is applicable for UHECR flux in general but not to separate events. If there exist another subdominant population of sources, it can produce some of the high energy events. In this case, we cannot exclude the origin of the TA highest energy particle from any of the source classes mentioned above.}. At the same time, our constraints on the distance to a closest source show that the source could be somewhere in a local galaxy group. This result agrees with other studies of general UHECR anisotropy~\cite{PierreAuger:2022axr, PierreAuger:2023mvf} and highest energy events~\cite{Fitoussi:2019dfe}. This encourages further searches of the closest source signature in various UHECR anisotropy observables. 

% \paragraph{Up to paragraphs.} We find that having more levels usually
% reduces the clarity of the article. Also, we strongly discourage the
% use of non-numbered sections (e.g.~\texttt{\textbackslash
%   subsubsection*}).  Please also consider the use of
% ``\texttt{\textbackslash texorpdfstring\{\}\{\}}'' to avoid warnings
% from the \texttt{hyperref} package when you have math in the section titles.

\acknowledgments

I am grateful to Peter Tinyakov for the general idea of this study and for the valuable remarks.
I also would like to thank Grigory Rubtsov and Sergey Troitsky for the fruitful discussions and comments and Oleg Kalashev for the help with the TransportCR code.
The work was supported by the Russian Science Foundation grant 22-12-00253. 

% \paragraph{Note added.} This is also a good position for notes added
% after the paper has been written.

% Bibliography

%% [A] Recommended: using JHEP.bst file
\bibliographystyle{JHEP}
\bibliography{biblio.bib}

%% or
%% [B] Manual formatting (see below)
%% (i) We suggest to always provide author, title and journal data or doi:
%% in short all the informations that clearly identify a document.
%% (ii) please avoid comments such as "For a review'', "For some examples",
%% "and references therein" or move them in the text. In general, please leave only references in the bibliography and move all
%% accessory text in footnotes.
%% (iii) Also, please have only one work for each \bibitem.

\end{document}